\documentstyle[epsfig]{mn}
\def\etal{{et al.}\thinspace}
\def\beb{\begin{thebibliography}{999}}
\def\eeb{\end{thebibliography}}
\def\bi{\bibitem{}}

\begin{document}
\title[Structure Formation and SZE]
{Constraints on structure formation models from the 
Sunyaev-Zel$^{\prime}$dovich Effect}

\author[S. Majumdar and R. Subrahmanyan]
       {Subhabrata Majumdar\thanks{E-mail : sum@physics.iisc.ernet.in}$^{1,2}$,
	    Ravi Subrahmanyan\thanks{ E-mail : Ravi.Subrahmanyan@atnf.csiro.au}$^{3,4}$ \\
$^1$ Joint Astrophysics Programme, Department of Physics,
Indian Institute of Science, Bangalore 560012, India \\
$^2$ Indian Institute of Astrophysics, Koramangala,
Bangalore 560034, India \\
$^3$ Australia Telescope National Facility, CSIRO, Locked bag 194, Narrabri,
	NSW 2390, Australia \\
$^4$ Raman Research Institute, Sadashivanagar, Bangalore 560080, India\\
} 

\maketitle

\begin{abstract}
In the context of cold dark matter (CDM) cosmological 
models, we have simulated images of the 
brightness temperature fluctuations in the cosmic microwave background (CMB) sky 
owing to the Sunyaev Zel$^{\prime}$dovich (S-Z) effect
in a cosmological distribution of clusters.
We compare the image statistics with recent ATCA limits on arcmin-scale 
CMB anisotropy.
The S-Z effect produces a generically non-Gaussian field and 
we compute the variance in the simulated temperature-anisotropy
images, after convolution with the ATCA beam pattern, for different 
cosmological models. 
All the models are normalised to the 4-year {\it COBE} data.
We find an increase in the simulated-sky temperature variance with increase in 
the cosmological density parameter $\Omega_0$.
A comparison with the upper limits on the sky variance set by the ATCA 
appears to rule out our closed-universe model: low-$\Omega_0$ open-universe 
models are preferred. The result is independent of any present day observations
of $\sigma_8$.

\end{abstract}

\section{Introduction}

A key problem in modern cosmology is the determination of 
the geometry of the universe. 
Classical methods of deriving the cosmological parameters involve measurements
of the redshift dependence of apparent luminosities of `standard candles', or
the angular sizes of `standard rulers', or the number densities of non-evolving
objects.  These methods are expected to probe the background geometry independent of any
structure formation.  Many of these classical approaches have been limited by
the difficulties in identifying objects, distributed over cosmic timescales,
that are untouched by astrophysical evolution; however, it may be noted that
progress has recently been reported using supernovae Type 1a as standard candles
(Perlmutter et al. 1998). 
 
The formation of structure is dependent on the background cosmology.
Attempts have been made to exploit this coupling by examining the parameter
space allowed for the cosmological constants by favoured models of structure formation.
Motivated by the discovery of 
anisotropy in the cosmic microwave background (CMB), progress in the understanding of
physical mechanisms responsible for the anisotropy spectrum and the
influence of the background cosmology in the generation of these anisotropies
have led to attempts at deriving constraints on the cosmological
parameters from the shape of the CMB anisotropy spectrum (Bond et al. 1994).
Anisotropies in the CMB are usually described by its
power spectrum: $C_{l}=\langle\mid C_{lm} \mid ^{2} \rangle$ 
represents the anisotropy power at multipole order $l$, where
$C_{lm}$ are the coefficients  of the spherical harmonic 
decomposition of the fractional temperature fluctuations. The  density 
fluctuations that are hypothesized to have been generated in the
very early universe 
are believed to grow via gravitational instabilities 
and give rise to the large scale structures we see in the present
day universe.  The CMB temperature fluctuations seen today on large angular scales
exceeding about a degree are believed to be a direct consequence
of matter inhomogeneities on scales exceeding $\approx 100$~Mpc
at the recombination epoch.
The gravitational and astrophysical evolution, in the post-recombination
universe, which led to the formation of galaxies and their clustering,
may have altered these primary radiation anisotropies and may have
given rise to the dominant  CMB fluctuations on small angular scales. 
Since the generation of the CMB anisotropies and their appearance
on the sky is intimately linked with the background cosmology,
the study of CMB anisotropies on different scales give constraints on 
theories of structure formation and the also on
the parameters of cosmological models. 

Anisotropies in the CMB are divided broadly into 
two classes: primary and secondary
anisotropies.  The primary anisotropies are 
those generated at the last scattering surface.
The angular power spectrum of these anisotropies, as seen by observers
at the present epoch, depends 
on the parameters of the background cosmology --- the Hubble constant $H_{\circ}$,
the mean matter density $\Omega_{\circ}$ and the constituents of the matter
({\it i.e.}, the amount of baryonic matter, cold dark matter, hot dark matter etc.)
and any cosmological constant $\Lambda$ --- and on the magnitude and spectrum
of primordial perturbations in the matter. 
Several authors have derived the marginal and joint dependencies of the
cosmological parameters on characteristic features in the CMB anisotropy 
spectrum. For example, at large and 
intermediate angular scales ($l \leq 500 $), for 
$\Omega_0 < 1$, the Doppler peaks in the CMB 
anisotropy spectrum are shifted towards larger $\l$ values
(see, for example, Kamionkowski \& Spergel 1994 or 
Lineweaver \& Barbosa 1998).  The ionization history modifies the
radiation spectrum and serves to attenuate power at high-order multipoles.
Primary anisotropies at small angular scales 
($l \geq 500$) are expected to be relatively damped in 
most models of structure formation  due to the thickness 
of the last scattering surface and the  diffusion damping of sub-horizon
scale baryon fluctuations in the pre-recombination era; 
anisotropies at even larger $\l$ 
may be critically dependent on the reionisation history.

The secondary anisotropies arise due to interactions between the 
growing matter perturbations and the CMB photons
as they travel from the last-scattering-surface to the present time. 
Arcmin-scale  anisotropies may be generated due to second-order mode coupling
between density perturbations and bulk velocities (Hu, Scott \& Silk 1994;
Persi et al. 1995).  Separately, decrements in the CMB sky may be generated
owing to the inverse-Compton scattering of 
background photons as they pass through concentrations of hot gas associated
with clusters of galaxies along the line of sight; this phenomenon is referred to 
as the Sunyaev Zel$^{\prime}$dovich (S-Z) effect.  
A dominant contribution to the anisotropy power at large $l$ may come from 
quasar-ionised hot gas bubbles (Aghanim et al. 1996).

Among the various effects that may give rise to 
CMB anisotropy on small angular scales, we focus on the S-Z effect in this work.
In section~2, we discuss some details of the S-Z effect and 
in section~3, we discuss some of the motivations
for attempting to constrain cosmological models using the S-Z effect.  
In section~4, we present the formalism we have adopted for deriving the expectations 
for CMB anisotropy on small angular scales for specific models 
of structure formation.  The results of simulations of sky temperature anisotropy
are compared with ATCA limits on arcmin-scale anisotropy to derive constraints
on the cosmological density parameter $\Omega_{\circ}$.  

\section{The Sunyaev-Zel$^{\prime}$dovich Effect}

The hot intracluster gas inverse-Compton scatters the CMB photons propagating 
through the cluster medium, and the energy transfer in this interaction between the
hot electrons and CMB photons results in a distortion to the CMB spectrum
(Sunyaev \& Zel$^{\prime}$dovich 1972).  
The integral of the electron pressure along any  line-of-sight 
through the cluster determines the magnitude of
the distortion in that direction, 
this is quantified in terms of the Compton $y$-parameter:

\begin{equation}
y=
\int {dl}{{k(T_{e}-T_{\gamma})}\over{m_{e}c^{2}}}n_{e}\sigma_{T},
\end{equation}

\noindent where $k$ is the Boltzmann's constant, $T_e$ and $T_\gamma$ are,
respectively, the kinetic temperature of the electron and the thermodynamic 
temperature of the incident black-body radiation, $m_e$ is the electron 
rest mass, $c$ is the velocity of light,
$n_e$ is the electron number density and $\sigma_T$ is the
Thomson scattering cross-section. 
 
The S-Z effect manifests itself as a change in sky brightness 

\begin{equation}
{\delta\textit{i}_\nu = \textit{y} \textit{j}_\nu(x), }
\end{equation}

\noindent towards the cluster with respect to the mean CMB intensity.
\textit{x} is a dimensionless frequency parameter defined to be

\begin{equation}
x = {{h\nu} \over {kT_o}},
\end{equation}

\noindent where $h$ is the Planck's constant, 
$\nu$ is the observing frequency and 
$T_{\circ}$ is the CMB temperature at the present epoch: $T_{\circ}=2.73$~K. 
The frequency dependence of the change in sky brightness, owing to 
the S-Z effect, is given by

\begin{equation}
\textit{j}_\nu(x) =
{{2(kT_{\circ})^3} \over {(hc)^2}}
 {{x^{4} e^{x}} \over {\left(e^{x} -1\right)}^{2}}
\left[{ {x}\over{\tanh\left(x/2\right)}}-4\right].
\end{equation}
 
\noindent The spectral function has a unique shape owing to the fact that although 
there is a net transfer of energy from the hot electron gas to the radiation,
inverse-Compton scattering conserves the total number of photons and the mean
photon energy is increased.  The result is that there is a net diffusion
of photons upwards in energy.
Assuming the cluster electron temperature
to be about 10~keV, the net distortion observed is expected to be zero 
at about 222~GHz (1.35~mm wavelength), 
the distortion is expected to result in a decrement in the
brightness temperature towards the cluster at lower frequencies and the cluster
will be seen as a positive source at higher frequencies.

At wavelengths longward of 1.35~mm, a cluster appears as a negative source on the
sky with respect to the mean CMB background intensity:
the flux density $S_\nu$ due to the integrated S-Z effect over the sky area 
of a cluster is

\begin{equation}
S_\nu(x) =
{{j_\nu(x)}\over{D_a^2(z)}} \int dV{{kT_e}\over{m_e c^2}} n_e \sigma_T.
\end{equation}

\noindent We have assumed that $T_e  \gg T_\gamma$. 
The integral extends over the cluster volume and $D_a(z)$ 
is the angular-size distance to the cluster.  
It is clear from the equation above that the net S-Z effect due to a cluster 
is  dependent only on the total 
mass of hot gas in the cluster and a density-weighted mean temperature. 
The S-Z flux density from a cluster diminishes as the square of its angular-size
distance: because the angular-size distance of objects at cosmological
distances, $z \ga 1$, saturates to a limiting value or even decreases with
increasing redshift, the S-Z flux from a cluster does not rapidly diminish with
increasing cosmological distance.  However, 
the distributions of cluster gas temperature and electron density may be functions
of redshift; therefore, the expected S-Z flux density from a cluster may
be redshift dependent.

It may be noted here that \textit{y} is a dimensionless parameter 
and \textit{$S_\nu$} 
has units of erg~s$^{-1}$~cm$^2$~Hz$^{-1}$: this flux density is 
usually expressed in Jansky (Jy), 
defined as $1~{\rm Jy} = 10^{-23}$ erg~s$^{-1}$~cm$^2$~Hz$^{-1}$.

\section {Motivation for studying the S-Z effect}

The Sunyaev-Zel$^{\prime}$dovich effect provides a rather nice 
and complementary approach to the traditional methods ---
which use X-ray temperature and X-ray luminosity evolution in
clusters of galaxies --- 
of studying the evolution of the mass function of collapsed objects
(Bartlett 1997).  The evolution in the abundance of clusters
of galaxies is sensitive to the mean matter density $\Omega_{\circ}$
and, consequently, is a useful constraint on cosmological models.
The traditional method derives its
faith from numerical simulations which show that 
there is a tight correlation between the virial
mass and the emission-weighted X-ray temperature in clusters
of galaxies (Evrard 1990, 1996).  However, the traditional method 
has a major disadvantage: because of `cosmological dimming',  the 
surface brightness of distant X-ray sources falls off as $(1 + z)^{-4}$,
and for this reason, obtaining samples of clusters at
cosmological distances can be observationally challenging.
In the case of the S-Z effect in cosmological clusters,  
the decrement in brightness in lines of sight towards
clusters of galaxies  has the distinct advantage of being independent
of the distance to the cluster.  
The S-Z flux density from a cluster will diminish 
with distance to the cluster
as the square of the angular-size distance; this is in contrast to X-ray
flux densities from clusters which diminish as the square of the luminosity
distance to the cluster.
Moreover, the integrated S-Z effect due to any cluster,
{\it i.e.,} the flux density decrement, is 
proportional to the total hot-gas mass times the particle-weighted temperature. 
Consequently, the detection of any  cluster is
independent of the gas' spatial distribution (assuming the  cluster is unresolved). 
If the observations resolve clusters, particularly at lower redshifts,
the observed sky temperature distribution will be sensitive to the 
temperature structure within the clusters; once again this may be contrasted
with X-ray emission images of cluster gas distributions which are sensitive
to the gas density distribution.

The S-Z effect may, therefore, be used as a tracer of
clusters and other massive hot gaseous objects at cosmological distances.
Because the redshift evolution of clusters is a sensitive probe of cosmology theory,
the number counts of S-Z sources may be used to infer 
their cosmological abundances and thereby deduce parameters relating
to the background cosmological model and the structure formation theory
(see, for example, Blanchard \& Bartlett 1998;
Oukbir \& Blanchard 1992, 1997).  Motivated by this reasoning, 
several authors have, in recent years, made 
calculations of the expected S-Z source counts and their 
redshift distributions.  These have been related to the 
cosmological parameters as well as the evolution
in the intra-cluster medium (Bartlett \& Silk 1994;
Barbosa et al. 1996; Colafransesco et al. 1997).

As discussed in section~2,
clusters of galaxies may be `visible', owing to the S-Z effect,
as sources in the sky with a negative flux density at wavelengths longward of 
1.35~mm and with a positive flux density at shorter wavelengths.
This S-Z effect has been imaged to-date towards  several clusters (see
Birkinshaw 1999 for a recent review).
We may consider distant cosmological clusters --- observed at wavelengths longward of 
1.35~mm --- as a 
sky distribution of sources with negative flux density. 
Observational sensitivity in modern radio telescopes is reaching
values close to that required for detecting the S-Z effect from cosmological
clusters towards `blank fields' where no obvious clusters are seen in either
their optical or X-ray emission.  
There have been claims in the literature of the detection of radio decrements (thought 
to be due to the S-Z effect) in sensitive images made of `blank' sky fields
(Jones et al. 1996; Richards et al. 1996); 
however, sensitive observations of several other 
fields with arcmin resolution 
have failed to detect any decrements or CMB anisotropies
(Subrahmanyan et al. 1998, 1999).  

It may be noted here that because the
S-Z effect has a generic non-Gaussian temperature distribution, it could be 
detected in a sky image by estimators sensitive to skewed variance.
Many other secondary contributors to the temperature 
anisotropy have 
a Gaussian distribution in amplitudes;
therefore, the negative-skewed-nature of the S-Z effect may be useful in distinguishing
it from other sources, foregrounds and instrument noise.

\section{Simulations}

\subsection{The cosmological distribution of clusters of galaxies} 

As discussed earlier, large-angle anisotropies in the  microwave background radiation 
may be traced back to its generation from small-amplitude
primordial density fluctuations in the early universe.
The matter perturbations 
have since grown due to gravitational instabilities
into the large scale structure we see at the present epoch.
In addition, as the primeval CMB radiation 
propagated to us through growing matter inhomogeneities, astrophysical couplings
give rise to secondary anisotropies on predominantly small angular scales. 
The power spectrum $P(k,z_{eq})$ of the fractional density fluctuations, 
at the redshift $z = z_{eq}$ when the energy densities in matter equalled
that in radiation, may be related to 
the primordial power spectrum $P_p(k)$ by:

\begin{equation}
P(k,z_{eq}) = T^2(k) \times P_p(k).
\end{equation}

\noindent The matter transfer function $T(k)$ describes the
processing of the initial density perturbations  
during the radiation dominated era (Padmanabhan 1993): the
modifications to the shape of the perturbations would depend on the
nature of the perturbations and the candidate dark matter. Thereafter,
during the matter-dominated era, the dominant dark matter (DM) perturbations
experience equal growth on all scales and 
$P(k,z)$ grows  retaining its shape.  The growth rate, described below by the
growth function, varies with cosmic time and 
depends on the mean matter density $\Omega_{\circ}$.
Models of structure formation are usually characterised
by specific $P(k,z_{eq})$ shapes and normalizations.

For the purpose of our simulations, we wish to relate the number density of collapsed
objects of different masses, at different cosmic epochs, to the initial density contrast
on different scales, {\it i.e.,} the initial matter spectrum $P(k,z_{eq})$.
It is assumed, in most theories of structure
formation, that the initial small-amplitude density fluctuations are Gaussian random,
{\it i.e.,} the amplitudes are Gaussian distributed and that the fluctuations on
different modes are uncorrelated (random phase fluctuations).  Structure is believed
to form from the growing perturbations hierarchically, with smaller-scale fluctuations
collapsing first and larger scales later.  The mass and redshift
distribution of the number density $n(M,z)$ of collapsed objects may then be computed 
using the Press-Schechter formalism (Press \& Schechter 1974; see Padmanabhan \&
Subramanian 1992 for a tutorial):
\begin{equation}
n(M,z)dM =
\sqrt{{2}\over {\pi}}
{{\langle \rho \rangle} \over {M}}{\nu_{c}}
\left| {{d{\ln {\sigma {(M,z)}}}}\over {d{\ln {M}}}} 
\right| {e^{-{\nu_{c}^{2}/2}}}{{dM}\over{M}},
\end{equation}
where
\begin{equation}
\nu_{c}(M,z)=
{{\delta_{c}}\over{\sigma (M,z)}} = 
{{\delta_{c} D_{g}(0)}\over{\sigma_{0}(M) D_{g}(z) }}.
\end{equation}
In these equations,
$\langle \rho \rangle$ denotes the mean co-moving matter density
in the universe: $\langle \rho \rangle = 3 \Omega_{\circ} H_{\circ}^2 / (8 \pi G)$,
where $\Omega_{\circ}$ is the density parameter, $H_{\circ}=100 h$~km~s$^{-1}$~Mpc$^{-1}$
is the Hubble constant and $G$ is the gravitational constant.
$\sigma(M,z)$ denotes the
rms fluctuations, at redshift $z$, in the fractional density contrast
in the matter when smoothed to a mass scale $M = {4 \over 3} \pi R^3 \langle \rho \rangle$,
where $R$ is the comoving radial length scale of the smoothing function.  
$\sigma_{\circ}(M)$ is the rms
density contrast at the present epoch.

The density contrast in an overdensity, computed using linear theory, at the epoch
when the collapsed object is deemed to have formed, is denoted by $\delta_c$.  We
adopt the collapse of a spherical top-hat overdensity  (Peebles 1980) as
a valid model for the description of the dynamical evolution of the peaks in the
Gaussian density distribution; the collapsed object is deemed to have 
`formed' at the cosmic time $t_c$, which is approximately twice the cosmic time
$t_m$ at which the overdense region attains maximum expansion radius.
In a universe with $\Omega_{\circ}=1.0$, $\delta_c = 1.68$.
It has been shown that 
$\delta_c$ varies by at most $\sim 4$ per cent for a range of
$\Omega_0$ from 0.1 to 1 (Lacey \& Cole 1993).  Therefore, we have chosen to adopt a constant value 
1.68 for $\delta_c$. 

In this Press-Schechter formalism, collapsed objects are identified
on the basis of their overdensity assuming linear growth of perturbations.
$D_{g}$, in equation~8, quantifies the growth factor
of the density perturbations from the
epoch of matter-radiation density equivalence ($z=z_{eq}$) to any later
epoch $z$ in the matter dominated era. 
In the absence of free streaming, 
the growth function is given by (Peebles 1980, Heath 1977)
\begin{equation}
D_g(z) = {{5\Omega_0}\over{2}}\left(1 + z_{eq}\right) g(z) 
       \int_{z}^{\infty} {{1 + z^{\prime}}\over {{g(z^{\prime})}^{3}}} dz^{\prime},
\end{equation}
where
\begin{equation}
g^2(z) = \Omega_0{(1 + z)}^{3} + \left( 1 - \Omega_0 - 
\Omega_{\Lambda}\right){(1 + z)}^{2}   + \Omega_{\Lambda}.
\end{equation} 
Closed form expressions are available in Weinberg (1972), 
Groth \& Peebles (1975) and  Edwards \&
Heath (1976) for universes with and without a cosmological constant $\Lambda$;
we have used these expressions for the purpose of our simulations.

The rms amplitude of the mass fluctuations at any redshift $z$: $\sigma(M,z)$, 
when smoothed with a spherically symmetric
window function of characteristic co-moving radius $R$, 
may be computed from the the matter power spectrum $P(k,z)$ using
the relation:

\begin{equation}
\sigma^{2}(M,z) =
 \int_{0}^{\infty} { {dk}\over{k} } { {k^{3}}\over{2\pi^{2}} } 
P(k,z){\left|\tilde{W_R}(k)\right|}^{2},
\end{equation}

\noindent where $\tilde{W_R}(k)$ is the Fourier transform of the corresponding  real 
space window function and, as before, $M = {4 \over 3} \pi R^3 \langle \rho \rangle$.
A spherical top-hat form with radius $R$ 
is usually adopted for the window function and this corresponds to a 
Fourier-space window function:
\begin{equation}
\tilde{W_R}(k)=
{{3}\over{{kR}^{3}}}{\left( \sin(kR) - kR\cos(kR)\right)}.
\end{equation}

The Press-Schechter formalism has been verified 
by Tozzi \& Governato (1998) against the results from numerical
n-body dynamical simulations, particularly for the mass range encompassing
clusters of galaxies. 

\subsection{P(k) and  its normalization}
 
\begin{figure*}
 \epsfbox[103 132 508 535]{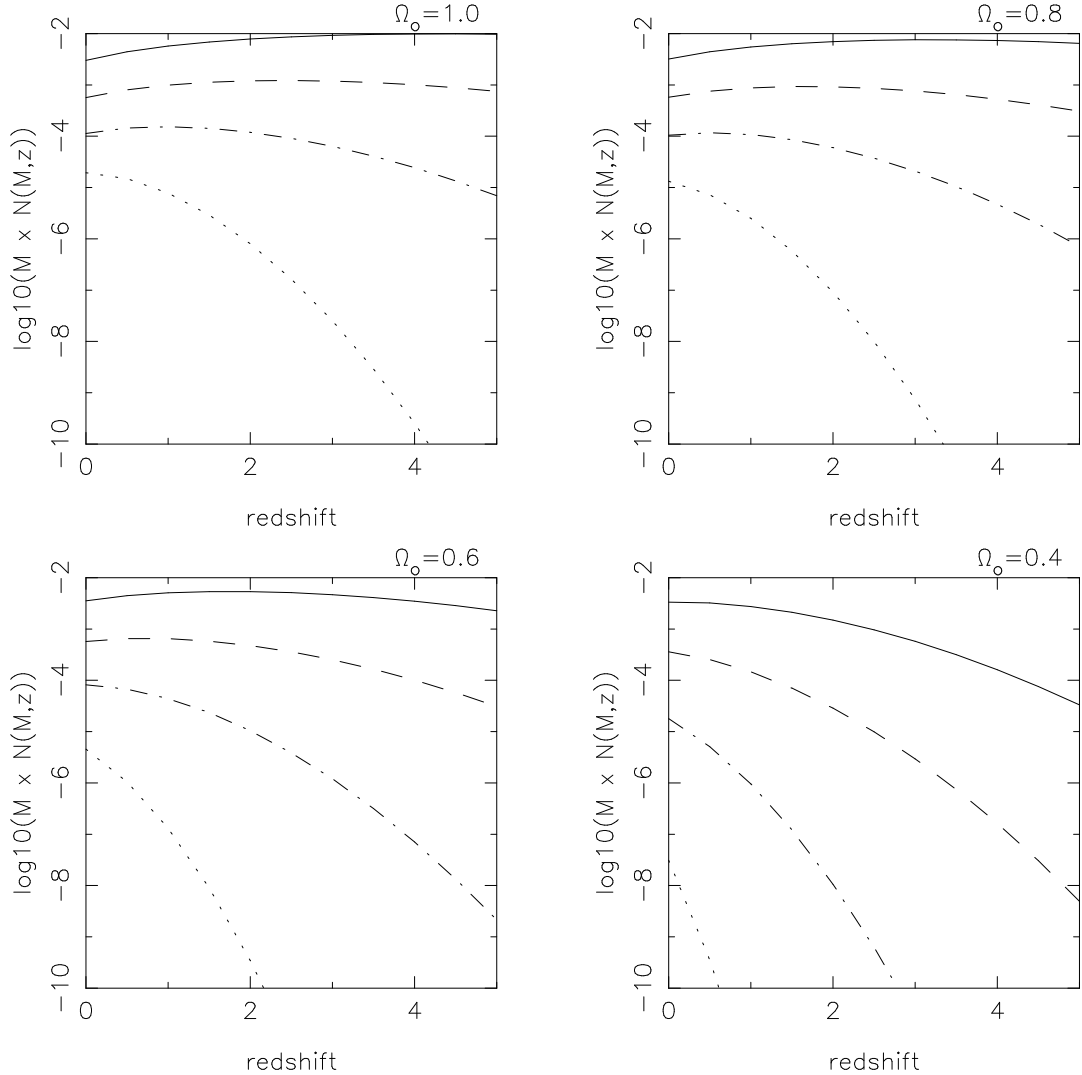}
 \caption{Abundance of objects with mass 
 	$M = 10^{12}$~M$_{\sun}$ (continuous line),
	$M = 10^{13}$~M$_{\sun}$ (dashed line),
	$M = 10^{14}$~M$_{\sun}$ (dot-dashed line) and
	$M = 10^{15}$~M$_{\sun}$ (dotted line).
	A CDM cosmology with $h=0.65$, $\Lambda=0$ 
	and $\Omega_{B}=0.019h^{-2}$ is adopted;
	the four plots are labelled with the assumed values of $\Omega_{\circ}$.
	The abundance $N(M,z)$ is expressed as the 
	number of objects per Mpc$^3$ and
	log$_{10}(M \times N(M,z))$ is plotted versus redshift $z$.}
\end{figure*}

We have adopted a power-law form for the initial primordial matter
spectrum with $P_{p}(k) \sim k^{n}$, where $n$ is the index of the 
primordial power spectrum; this index $n$ 
equals unity for Harrison-Zel'dovich scale-invariant spectra.
The power spectrum at any redshift $z$ in the matter dominated era may then be
written in the form

\begin{equation}
{{k^{3}}\over{2\pi^{2}}}P(k,z)=
{\left({{ck}\over{H_0}}\right)}^{3+n} 
\delta^{2}_{H} T^{2}(k) D^{2}_{g}(z)/D^{2}_{g}(0),
\end{equation}

\noindent where the $D_g$'s are the growth factors defined in section~4.1.
$\delta_H$ represents the amplitude normalization 
of $P(k,z)$ and is defined as the amplitude of perturbations on the 
horizon scale at the present epoch $z=0$. 

The large-angle CMB anisotropies discovered by $COBE$ (Smoot et al. 1992)
are believed to have been generated by processes at the last scattering surface.
Precisely for the reason that the matter fluctuations were
of small fractional amplitudes at the epoch $t$ they generated these anisotropies,
and because the scale length of the perturbation modes well exceeded $ct$,
the coupling physics between the matter perturbations and radiation was 
linear and did not involve any astrophysical interactions.  The CMB anisotropies
detected by $COBE$ may, therefore, be related to the primordial matter spectrum
fairly exactly without all the uncertainties
associated with the astrophysics involved in the formation of collapsed baryonic
objects at late epochs.  For this reason, we have chosen to normalize
the $P(k)$ with the $COBE$ results.

We have used the 
fitting formulae of Bunn \& White (1997) to normalize 
the power spectrum to the $COBE$-$DMR$
measurements.  For an open universe with vanishing cosmological constant,
with no CMB anisotropies coming from gravitational
waves, we use the normalization

\begin{equation}
\delta_H= 1.95\times{10}^{-5}\Omega_0^{-0.35 -0.19\ln{\Omega_0} - 0.17\tilde{n}}
                                          e^{-\tilde{n}- 0.14{\tilde{n}}^{2}}.
\end{equation}

\noindent $\tilde{n} = n - 1$ and the normalization is 
valid for $0.7 \leq n \leq 1.2$.  The fit is also valid over the 
range 0.2--1.0 in $\Omega_0$. 

As mentioned earlier, the shape of $P(k,z)$ in the matter dominated era is
completely determined by the transfer function $T(k)$ and the assumed form for
the primordial spectrum.  With the choice of a power-law form for $P_p(k)$,
the transfer function is critical in relating the power on 
small wavenumber modes  --- which is 
fixed by the adopted $COBE$ normalization --- to 
those on large $k$ modes at which we have clusters of galaxies.  There are
several fitting forms available in the literature for $T(k)$ in the context of the
cold dark matter (CDM) dominated universes.
However, different parameterizations of
transfer functions can differ by large amounts (Peacock \& Dodds 1993) and,
moreover, to obtain
better accuracy the effect of baryon damping must be included
(Hu \& Sugiyama 1996; Ma 1996;
Eisenstein \& Hu 1999). We, however, work within the context of 
low-baryon-density universes and, therefore, include the
effect of non-zero baryon content by adopting a modified `shape factor' $\Gamma$.
In this work,  we have adopted
the fitting function provided by Bardeen et al. (1986):
\[
T_{CDM}(q) = {{\ln\left(1 + 2.34q\right)}\over{2.34q}} \times
\]
\begin{equation}
~~~~~~~~~~{\left[ 1 + 3.89q + (16.1q)^{2} + (5.46q)^{3} + 
	(6.71q)^{4} \right]}^{-1/4} 
\end{equation}

\noindent where  $q=k/h\Gamma$ and 
$\Gamma=\Omega_0h\exp\left(-\Omega_B -\Omega_B/\Omega_0\right)$. 
$\Omega_B$ is the baryon density parameter and is the ratio of the mean density
of baryons in the universe to the critical density $\rho_c = 3H_{\circ}^2/(8 \pi G)$.

The Press-Schechter formalism described in section 4.1, along with the
$COBE$-normalized $P(k)$ defined above, have been used to calculate the
number densities of collapsed objects. We choose to identify the condensates as 
groups/clusters of galaxies and have computed their abundances 
as a function of cluster mass and redshift for a set of 
CDM cosmologies differing in their $\Omega_{\circ}$. 
We have not considered the effect of including a cosmological constant in this work 
and have put $\Lambda=0$.  The 
computations of cluster distributions $n(M,z)$ have been carried out 
over logarithmic bins in the  mass range $10^{13}$--$10^{16}M_{\odot}$. 
The distribution in mass is computed at redshifts
spaced at intervals of $\Delta z = 0.1$.  Plots of $n(M,z)$ distributions 
are shown in Fig.~1.  

The theory predicts that the
cosmic abundance of massive collapsed objects are extremely sensitive 
to the amplitude and slope of the 
primordial power spectrum and also on the growth factor.  
In a flat cosmology with $\Omega_{\circ} = 1$,
$P(k)$ would grow as $(1 + z)^{-2}$; in an open universe the growth approximately
follows this evolution down to redshift $z \approx (\Omega_{\circ}^{-1} - 1)$
and is stunted thereafter. Therefore, in these models, 
which are all $COBE$ normalized, the total growth up to the present time 
will be greater in 
models with larger $\Omega_{\circ}$.  Consistent with this
reasoning, it may be seen in Fig.~1 that the larger $\Omega_{\circ}$ models 
are dynamically more evolved: they have larger numbers of $10^{15}$~M$_{\sun}$
objects at $z=0$ and in these models the abundances of lower mass objects are
declining with cosmic time as they are incorporated into larger mass objects.

It may be seen from Fig.~1 that the decline, with redshift, 
in the abundance of high-mass ($10^{15}$~M$_{\sun}$)
collapsed objects relative to the abundance at $z=0$ is steepest in the case of
models with lower $\Omega_{\circ}$.  This is the opposite of the expectations
for models which are normalized at the present time to, for example, a specific
value of $\sigma_8$, which is the rms mass fluctuations when the present day
universe is smoothed using a window function with a radius $R = 8h^{-1}$~Mpc.
Although the growth function evolves slower at low redshifts in models with
low $\Omega_{\circ}$, the abundance of massive collapsed objects evolves
more strongly in the case of the low $\Omega_{\circ}$ models because in these
models the rms mass fluctuations $\sigma(M,z)$ is itself low at $z=0$.

The Gaussian characteristic of the Press-Schechter mass 
function is evident in the distributions: at any redshift, 
there is a rapid exponential decline in the number 
density of objects in larger mass bins.

\subsection{Simulated sky images} 

Our simulation codes compute the abundances $n(M,z)$ in redshift slices
in the redshift range $z=0$--5.  We make the conservative assumption that
collapsed objects in the restricted range
$10^{13}$--$10^{16}~M_{\sun}$ alone represent clusters of
galaxies containing hot gas.  
Adopting a model, described below, for the spatial
distribution of hot gas in the potential wells of these clusters,
we simulate sky images of the S-Z effect expected owing to these
clusters.  The sky was simulated by separately computing the contributions from
different redshift slices along the line of sight.
In any redshift slice,  
the clusters are assumed to be Poisson-random distributed on the sky.
We have simulated
square patches of sky consisting of 256 pixels along each side: the pixels were 
chosen to be 10~arcsec square making the total image size
$42^{\prime}40^{\prime\prime}$.  The redshift slices were of size $\Delta z = 0.1$.

The mean number of clusters, $\lambda$,
with mass in the range $M$--$(M+\delta M)$, is related to the 
Press-Schechter number density by:
\begin{equation}
{\lambda} = n(M,z) \times \delta M \times {\rm volume~corresponding~to~a~pixel}.
\end{equation} 
The comoving volume of any pixel in a slice at redshift $z$
is calculated as the product of the comoving area $(\Delta l)^2$ covered by
the square pixels of angular size $\Delta \theta = 10$~arcsec 
and the comoving line-of-sight
distance $\Delta s$ corresponding to the redshift slice $\Delta z$.
These are given by:
\begin{equation}
\Delta s = 
{{c}\over{H_0}} 
	{ {dz} \over { (1+z) \sqrt{1+\Omega_{\circ} z} }},
\end{equation}
and
\begin{equation}
{\Delta l} = { {2 c \Delta \theta} \over {H_{\circ} \Omega_{\circ}^2 (1+z)} }
	\left\lbrace \Omega_{\circ}z + (\Omega_{\circ}-2)
	\left\lbrack \sqrt{\Omega_{\circ}+1} - 1 \right\rbrack \right\rbrace.
\end{equation}

We have associated every collapsed massive object, 
with mass exceeding $10^{13}~M_{\sun}$,
with cluster gas.  In order to model the spatial distribution of the intra-cluster
gas, the isothermal $\beta$ model has been adopted, 
in which the cluster gas is modelled as
being spherically symmetrical, centred in the gravitational potential of the cluster 
and shock heated to a temperature corresponding to the infall energy. 
The variation in gas density with radial distance $r$ is assumed to be given by

\begin{equation}
\rho (r) = \rho_{0}
{[1 + {(r/r_c)}^{2}]}^{-3\beta /2},
\end{equation}
 
\noindent where $\rho_{\circ}$ is the central density and $r_c$ is the core radius.
The value of $\beta$, inferred from the X-ray surface-brightness profiles observed in
clusters of galaxies,  is believed to be in the range
0.5--0.9 (Markevitch et al. 1998, Jones \& Forman 1984). Herein, for simplicity,  we 
adopt a value of ${2\over3}$ for $\beta$.

Assuming that the dynamical 
collapse of the cluster is self similar, the scaling of the 
core radius $r_c$ of collapsed objects 
with mass and redshift  has been derived to have the form
(Colafrancesco 1997, Kaiser 1986): 
\[
r_c(\Omega_0,M,z) = {{1.3 h^{-1} {\rm Mpc}}\over{p}} {{1}\over{(1 + z)}}~ \times 
\]
\begin{equation}
~~~~~~~~~~{\left[{{M}\over{10^{15}h^{-1}M_{\odot}}} {{\Delta\left(\Omega_{0} = 1, z = 0\right)}
\over{\Omega_0 \Delta(\Omega_0,z)}}\right]}^{1/3}.    
\end{equation}
In this parametrization, the non-linear density contrast at the epoch of  
virialisation  is $\Delta = \tilde{\rho}/\rho$ where
$\tilde{\rho}$ is the density contrast and $\rho$ is the mean background density.
For the case of an  open universe, 
\begin{equation}
\Delta = {{18\pi^{2}}\over{\Omega_0 H^2_0 t^2_v}} {{1}\over{{\left(1 + z_v\right)}^{3}}},
\end{equation}
where $t_v$ and $z_v$ are the cosmic time and redshift corresponding to
the epoch of virialisation.  In the case of a flat universe,
$\Delta \approx 178$ (Peebles 1980; Colafransesco et al. 1997).
In the above parametrization, $p$ is an adjustable free parameter that may be selected to
fit observations of nearby clusters.  If $r_m$ denotes the maximum radial extent of the cluster,
$p=(r_m/r_c)$ and is a measure of the concentration of the cluster mass: small values of $p$
model the cluster gas as centrally concentrated where as larger values of $p$ spread the
gas mass away from the core.  Values of $p \approx 6$
appear to better model the parameters of nearby clusters; a comparison of the model 
with the observed parameters for the Coma cluster is given below.
We propose to use the
simulations for the purpose of predicting the sky variance as observed by the ATCA; these
observations are made with arcmin resolution and will be less sensitive to S-Z effects
from clusters whose gas distribution is extended.  The ATCA observations
couple somewhat better to models with smaller $p$; therefore,
we have adopted the conservative choice of $p=8$.  It may be noted that sensitive
X-ray imaging of the gas in nearby clusters have shown
gas extending out to at least eight core radii.

Although recent observations indicate that intracluster gas has temperature structure 
(Markevitch et al. 1998), we have modelled the clusters as being isothermal.
It follows from the assumption that clusters form from self-similar
collapse that the temperature 
$T \propto {{M}\over{R}}$ and a good approximation for the temperature of the
intra-cluster gas in clusters is (Colafrancesco et al 1997)
\[
T = 6.7\times 10^{7} (1 + z)
	{\left[{{M}\over{10^{15}h^{-1}M_{\odot}}}\right]}^{2/3}~\times
\]
\begin{equation}
~~~~~~~~~~~~{\left[{\Omega_0 \Delta(\Omega_0,z)}\over 
         {\Delta\left(\Omega_{0} = 1, z = 0\right)}\right]}^{1/3}~~{\rm K}.
\end{equation}
The relation between the mass and the temperature in equation(22) agrees well
with the recent M-T parametrisation  as suggested by Bryan \& Norman (1998).
We have assumed that the hot gas detected in its X-ray emission is responsible for
any S-Z effect as well (Colafrancesco et al. 1994; Blanchard et al. 1992;
Vittorio et al. 1997; Kaiser 1986).

Only the baryonic matter in collapsed clusters gives the S-Z effect.  White et al. (1993)
have estimated the baryon content in clusters to be 
\begin{equation}
{{M_b}\over{M_{tot}}} \geq 0.009 + 0.050h^{-3/2},
\end{equation}
where the first contribution is due to the galaxies and the second is due to the 
intra-cluster gas.  Primordial nucleosynthesis predicts a universal baryon abundance
$\Omega_b \simeq 0.019h^{-2}$ (Burles et al. 1999).  Adopting the Hubble parameter $h=0.65$,
it is seen that the baryon gas-mass fraction in nearby clusters is about 0.1 and is a
factor of two greater than the universal baryon abundance.  The $Einstein$ Medium Sensitivity
Survey (EMSS) data appears to indicate a decline in the abundance of X-ray luminous clusters
with redshift (Henry et al. 1992); this may be parametrized as a hot-gas fraction that
evolves as $t^{1.4}$ with cosmic epoch.  It may be noted that adopting a gas fraction
that declines with redshift is a conservative assumption because it reduces the predicted
S-Z effect.  We have adopted a parametrization
\begin{equation}
{{M_{gas}}\over{M_{tot}}} = 0.050h^{-3/2} 
\left\lbrack {M_{tot}\over{10^{15}h^{-1}~{\rm M}_{\sun}}} \right\rbrack ^{0.2} 
\left\lbrack {t \over {t_{\circ}}} \right\rbrack ^{q},
\end{equation}
with $q=1.4$.  $M_{gas}$ represents the hot gas mass in the cluster and $M_{tot}$
represents the total mass in the collapsed object. The exponent $0.2$ in
equation(24) is from a fit in Colafrancesco et al (1997).
We have assumed that the intra-cluster gas has a primordial composition, with 
helium and hydrogen atoms in the number ratio 1:10, and is fully ionized.  

The cluster mass distribution, as also the intra-cluster gas, are modelled as truncated
at a maximum radius $r_{m} = p\times r_c$.  We define the `impact parameter' $b$ 
as the projected distance between the
cluster centre and any line of sight through the cluster.
It follows that the S-Z decrement in the Rayleigh-Jeans
portion of the spectrum, in units of temperature, expected at any sky position 
$b \la r_{max}$ is given by
\begin{equation}
\Delta T = -T_{CMB} {{2kT}\over{m_e c^2}} \times \int_b{n_e \sigma_{T} dl},
\end{equation}
where $T_{CMB}$ is the average brightness temperature of the CMB.  
The integral is computed along the line of sight
through the cluster at the impact parameter $b$.  

The assumption that the cluster gas is of primordial composition implies that the
total number of hot electrons in the cluster is 
$N_e = (18/14)(M_{gas}/a.m.u)$.  The S-Z decrement may be written in the form
\[
\Delta T/T_{CMB} = - {{2kT}\over{m_e c^2}} \sigma_T~~\times
\]
\begin{equation}
~~~~~~~~~ {{1}\over{4\pi}}
{{\tan^{-1}({{\sqrt{r_{max}^{2} - b^{2}}}\over{\sqrt{r_c^2 + b^2}}})N_e r_c} \over
{\sqrt{r_c^2 + b^2}[r_c r_{max} - r_c^2 \tan^{-1}({{r_{max}}\over{r_c}})]}}.
\end{equation}

The Coma cluster at a redshift 0.0235, with a total mass about 
$2 \times 10^{15}$~M$_{\sun}$ , is observed to have X-ray gas with a temperature 8--9~keV and
a core radial size of 10.5 arcmin (Silverberg et al. 1997, Herbig et al. 1995).  
The Compton-$y$ parameter has
been measured to be $9 \times 10^{-5}$ towards its centre (Herbig et al. 1995).
In our simulations,
for the choice $p=6$ and adopting a value of 0.096 as the baryon gas fraction,
we find that a collapsed object of mass $2 \times 10^{15}$~M$_{\sun}$ at redshift
$z=0.0235$ --- corresponding to the Coma cluster parameters --- yields values of
10.3--11.0~arcmin for the core size, 8--9$\times10^7$~K for the 
gas temperature and the central S-Z Compton-$y$ decrement 
is in the range 7--8.5$\times10^{-5}$.

We have accumulated the S-Z effect from clusters in redshift slices up to a maximum
redshift of 5.  In each redshift slice, the pixels were populated by collapsed point masses
in a Poisson random fashion; the expectation that any pixel was populated by
objects in any mass bin was governed by equation~16.  We then substituted our cluster
gas model for every mass point: the S-Z effect owing to each cluster is distributed
over several pixels surrounding that at which the point mass was located.
The variance of the temperature fluctuations in the 
cumulative S-Z effect images were computed.  

\begin{figure*}
 \epsfbox[112 140 511 534]{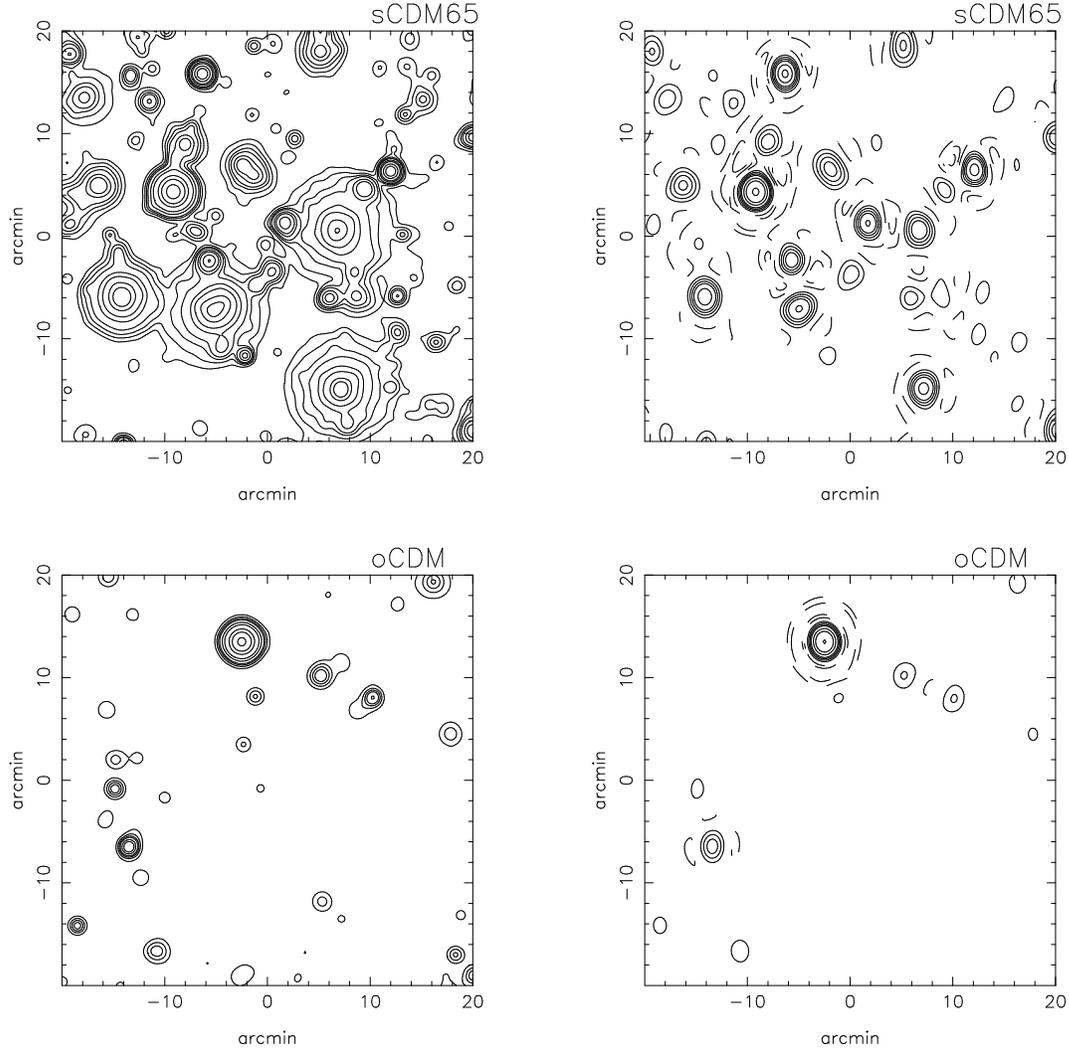}
 \caption{ Sample images.  The upper pair of images corresponds to CDM cosmology
	with $\Omega_{\circ}=1.0$; the lower pair corresponds to 
	$\Omega_{\circ}=0.6$. The cluster gas distribution has been parametrized with
	$p=8$ and the baryon gas fraction is assumed to evolve with redshift as described by
	equation~24; the Hubble parameter $h=0.65$ has been adopted.  The images on the
	left have been convolved with a Gaussian beam of FWHM 1~arcmin; the images on 
	the right have been convolved by the ATCA product beam.  Contours at 
	($-$3, $-$2, $-$1, 1, 2, 3, 4, 6, 8, 12, 16, 24)~$\times~-30~\mu$K for the
	images on the left and $-30~\mu$Jy~beam$^{-1}$ for the images on the right.}
\end{figure*}

We have compared the results of the simulations with the observational limits on
arcmin-scale CMB anisotropy set by the ATCA experiment at 8.7~GHz (Subrahmanyan et al. 1993, 
1998, 1999).  These observations have been made with a Fourier synthesis array.  The
sky is viewed by the telescope as attenuated by a primary beam that has a full width
at half maximum of about 6~arcmin and the synthetic images represent the convolution
of the visible sky with a synthetic beam pattern.  The ATCA experiment sets upper limits
on any residual sky fluctuations apart from the instrument noise and confusing foreground
sources.  Assuming `flat-band' CMB fluctuations, the ATCA limit corresponds to 
$Q_{flat} \la 25~\mu$K and, for their telescope filter function,  
this corresponds to an upper limit of 21~$\mu$Jy~beam$^{-1}$
on the flux-density fluctuations on the sky.
We have convolved the images resulting from our simulations by a beam that is the
product of the ATCA synthetic beam and primary beam.
Sample images are shown in Fig.~2 corresponding to 
CDM models with $\Omega_{\circ}=1.0$ and 0.6.  Images are displayed using a 1-arcmin 
Gaussian (normalized to give unity volume) 
as the convolving beam and separately using the ATCA product beam (normalized to have
a peak unity) as the convolving function.  It is seen that the extended clusters, seen in
the $\Omega_{\circ}=1.0$ simulation, are resolved by the ATCA beam: the extended
S-Z structures will not be detectable by the ATCA imaging.  
It may be noted that before the
convolution, the image pixels are in units of temperature (K);
following the convolution, the pixel intensities are in units of 
flux density (Jy~beam$^{-1}$).  The predictions for the image variance
have been made for different plausible cosmological models.

\section{Results and Discussion}

\begin{table*}
\begin{minipage}{160mm}
\caption{Statistics for different cosmological models.}
\begin{tabular}{cccccccc}
Model & $\Omega_0$  & $h$ & $\sigma_8$ & gas\,fraction\footnote{The 
parametrization of the gas fraction is described in equation~24.}
& mean $\,y$ & Image rms
& Image rms\\
&&&&& $\times 10^{-5}$ & before \,convolution &  after \,convolution \\
&&&&&& ($\mu$K) & ($\mu$Jy~beam$^{-1}$) \\
sCDM           & 1.0 & 0.50  & 1.128 & parametrized & 0.61 &  81 & 28 \\
sCDM(high $h$) & 1.0 & 0.65 & 1.567 & parametrized & 1.9  & 313 & 58 \\
oCDM1          & 0.8 & 0.65 & 1.326 & parametrized & 0.46 &  60 & 23 \\
oCDM2          & 0.6 & 0.65 & 1.004 & parametrized & 0.18 &  38 & 15 \\
oCDM3          & 0.4 & 0.65 & 0.605 & parametrized & 0.05 &   9 &  3 \\
sCDM(variant1) & 1.0 & 0.50 & 1.128 & no evolution & 1.9  & 207 & 72 \\
sCDM(variant2) & 1.0 & 0.50 & 1.128 & ${{M_{gas}}\over{M_{tot}}}=\Omega_B$ &
2.4 & 245 & 79 \\
sCDM(p=6)      & 1.0 & 0.50 & 1.128 & parametrized & 1.16 & 98 & 37 \\

\end{tabular}
\end{minipage}
\end{table*}

Results of the simulations are given in Table~1 for a range of model parameters.  The
image variances have been listed both before and after convolving with the ATCA product
beam.  The standard CDM model with $h=0.5$, $\Omega_{\circ}=1.0$ and with the hot gas
fraction parametrized as described in equation~24, is expected to result in an image rms 
which exceeds the ATCA limit.  Increasing the Hubble parameter to a more likely value of
$h=0.65$ increases the expectations for the image rms and this `high-$h$' CDM model is
rejected with greater confidence.  Considering open universe CDM models with 
$\Omega_{\circ} < 1$, the expected image variance decreases with decreasing $\Omega_{\circ}$.
Low $\Omega_{\circ} < 0.8 $ open-universe models are allowed by the ATCA limits.

If we assume that $q=0$ and that the cluster gas fraction does not diminish with redshift,
high-redshift clusters contribute to the net S-Z sky fluctuations.  This is seen in the table
where the image rms rises from 28 to 72 $\mu$Jy~beam$^{-1}$ when the no-evolution assumption
is made for the standard CDM model.  We have also simulated images corresponding to the
case where the gas fraction is a constant and equal to the mean baryon
density predicted by nucleosynthesis.  This gas model leads to predicted rms values which
are not very different from that given by the $q=0$ parametrized model; this indicates that
the lower mass ($M_{tot} \approx 10^{13-14}~$~M$_{\sun}$) objects dominate the variance
contribution in the no-evolution case.

Simulations of images corresponding
to the choice $p=6$ leads to enhancements in the expectations of the
image rms values confirming that our earlier choice ($p=8$) was relatively conservative.

We have computed the mean Compton-$y$ parameter from the images for the different models
and these are also listed in Table~1.
The $COBE$-$FIRAS$ experiment has set an upper limit of $y < 1.5 \times 10^{-5}$ 
(Fixsen et al.
1996).  It may be noted that our sCDM model  with $h=0.65$ is independently 
rejected by these limits.  Likewise, the models with no evolution in gas fraction are
also disallowed by the $FIRAS$ results.

The value of $\sigma_8$ has also been computed for each model and listed in Table~1.  
The standard CDM models
normalized to $COBE$ are known to be incompatible with estimates of $\sigma_8$
derived from the local X-ray luminosity function or other local measures of galaxy
clustering (which prefer low values of $\sigma_8$ about 0.6).  
The ATCA results independently
provide evidence in favour of a low-$\Omega_{\circ}$ universe. It must be
emphasised once again that the sky fluctuations due to SZE is generically
non-gaussian in nature and a precise elimination of models have to take this
into account.

An important difference between this work as compared to previous predictions for CMB
anisotropies from S-Z effects is that we have normalized our matter power spectrum to 
the $COBE$ anisotropy results.  All the previous predictions that we are aware of 
have normalized their models to observations of the present day clustering in galaxies
or to X-ray luminosity functions;
consequently, other workers have essentially normalised their matter power
spectra to $\sigma_8$. This makes it possible for all the models to be
consistent with the observational estimates of present day cluster abundances.
Our choice of $COBE$ normalisation results in a $\sigma_8$ that varies across
the models. This results in a dependance on $\Omega_{\circ}$ which is opposite
of what usually found. The difference may be understood in the following way :
if the $\sigma_8$ is held constant across cosmological models, as previous
workers have done, varying $\Omega_{\circ}$ changes the growth function and
consequently the abundance of clusters at redsifts $z > 0$ will decrease with
increasing $\Omega_{\circ}$. On the other hand, when the matter power spectrum
is normalised to $COBE$-$DMR$ data, varying $\Omega_{\circ}$ across models alters
the shape of the matter spectrum and the normalisation in addition to the growth
function. In this case, increasing $\Omega_{\circ}$ results in an increase in
cluster abundances at all redshifts.
Our choice of normalization
results in a $\sigma_8$ that varies across the models.

\section{Conclusion}
We have considered cosmological models composed of cold dark matter and baryons (and no
cosmological constant), having an initial scale-invariant spectrum of
adiabatic perturbations. We have used the Press-Schechter formalism to generate the 
distribution of
clusters. We have normalised the rms mass fluctuations to $COBE-DMR$ data. 
The Sunyaev-Zel$^{\prime}$dovich decrement is calculated for each cluster adopting a model for
the density profile of the cluster gas. We have simulated blank sky surveys
for the S-Z effect owing to a cosmological distribution of clusters.  
Finally, we have predicted the expectations for the variance in the background sky
both before and after convolving with the ATCA beam. Based on a comparison with the 
upper limits set by the ATCA on CMB anisotropies on arcmin scales,
we conclude that $COBE$-normalized CDM models of the universe 
with a high density parameter ($\Omega_{\circ} > 0.8$) are rejected.  This result is
independent of any present epoch measures of $\sigma_8$.

\section*{ACKNOWLEDGEMENTS}      
 SM would like to thank Dr. Pijushpani Bhattacharjee
for his constant encouragement, Dr Biman Nath for many valuable discussions and
Dipanjan Mitra for helping with the computers. SM also wishes to thank
``everyone@rri'' for their hospitality and warmth and for providing access
to the computer facility.

\beb

\bi Aghanim N., Desert F.~X., Puget J.~L., Gispert R., 1996, A\&A, 311. 1
\bi Barbosa D., Bartlett J.~G., Blanchard A., Oukbir J., 1997, 
in : Microwave 
Background Anisotropies, Proceedings of XIV Moriond Astrophysics Meetings, ed: Bouchet
F.,
Gispert R., Guiderdoni B., Jean Tran Thanh Van
\bi Bartlett J.~G., Silk J., 1994, ApJ, 423, 12
\bi Bartlett J.~G., 1997 in: From Quantum Fluctuations to Cosmological Structures,
	School held
in Casablanca, ed Valls-Gabaud D., Hendry M.~A., Molaro P., Chamcham K., A.S.P Conf 
Series, vol 126,
\bi Birkinshaw M., 1999, Phy Rep, 310, 97
\bi Blanchard A., Valls-Gabaud D., Mamon D.~A., 1992, A\&A, 264, 365
\bi Blanchard A., Bartlett J.~G., 1998, A \& A, 332, 49L
\bi Bond J.~R.,Crittenden R., Davis R.~L., Efstathiou G., 
	Steinhardt P.~J., 1994, PRL, 72, 13
\bi Bryan G.~L., Norman J.~L., 1998, ApJ, 495, 80	
\bi Bunn E.~F., White D., 1997, ApJ, 480, 6
\bi Burles S., Nottel K.~M., Turner M.~S., astro-ph/9903300
\bi Colafrancesco S., Mazzotta P., Rephaeli Y., Vittorio N., 1994, ApJ, 433, 454
\bi Colafrancesco S., Mazzotta P., Rephaeli Y., Vittorio N., 1997, ApJ, 479, 1
\bi Edwards D., Heath D., 1976, Ap Space Science, 41, 183
\bi Eisenstein E.~J., Hu W., 1999, ApJ, 511, 5
\bi Evrard A.~E., 1990, in: Clusters of Galaxies, ed: Oegerle W.~R., Fitchett M.~J., Danly L, 
	Cambridge University Press, New York
\bi Evrard A.~E., Metzler C.~A., Navarro J.~F., 1996, ApJ,  469, 494
\bi Fixsen \etal, 1996, ApJ, 473, 576
\bi Groth E.~J., Peeble P.~J.~E., 1975, A\&A, 41, 143
\bi Heath D.~J., 1977, MNRAS, 179,351
\bi Henry J.~P. \etal 1992, ApJ, 386, 408
\bi Herbig T., Lawrence C.~R., Redhead A.~C.~S., Gulkis S., 1995, ApJ, 449, L5
\bi Hu W., Scott D., Silk J., 1994, Phys. Rev. D. 49, 2
\bi Hu W., Sugiyama N., 1996, ApJ, 471, 54
\bi Jones C., Forman W., 1984, ApJ, 276, 38
\bi Jones M., \etal, 1997, ApJ, 479, L1
\bi Kaiser N., 1986, MNRAS, 222, 323
\bi Kamionkowski M., Spergel D.~L., 1994, ApJ, 432,7
\bi Lacey C., Cole S., 1993, MNRAS, 262, 627
\bi Lineweaver C.~H., Barbosa D., 1998 ApJ, 496, 624
\bi Ma C.~P., 1996, ApJ, 471, 13
\bi Markevitch M., Forman W.~R., Sarazin C.~L., Vikhlinin A., 1998, ApJ, 504, 27
\bi Oukbir J.,  Blanchard A., 1992, A\&A, 262, L21
\bi Oukbir J., Blanchard A., 1997, A\&A, 317, 1
\bi Padmanabhan T., 1993 Structure Formation in the Universe, Cambridge University Press,
Cambridge
\bi Padmanabhan T., Subramanian K., 1992, Bull. astr. Soc. India, 20, 1
\bi Peacock J.~A., Dodds S.~J., 1994, MNRAS, 267, 1020
\bi Peebles P.~J.~E., 1980, Large Scale Structure of the Universe., Princeton 
University Press, Princeton
\bi Perlmutter S., \etal 1998, Nature, 391, 51
\bi Persi F.~M., Spergel D.~N., Cen R., Ostriker J.~P., 1995, ApJ, 442, 1
\bi Press W.~H., Schecter P., 1974, ApJ, 187, 225
\bi Richards E.~A,, Fomalont E.~B., Kellermann K.~I., Partridge R.~B., Windhorst
 R.~A., 1997,
 AJ, 113, 1475
\bi Silverberg R.~F., \etal, 1997, ApJ, 485, 22
\bi Smoot G.~F.,\etal, 1992, ApJ, 396, L1
\bi Subrahmanyan R., Ekers R.~D., Sinclair M., Silk J., 1993, MNRAS, 263, 416
\bi Subrahmanyan R., Kesteven M.~J., Ekers R.~D., Sinclair M., Silk J., 1998,
	MNRAS, 298, 1189 
\bi Subrahmanyan R., Kesteven M.~J., Ekers R.~D., Sinclair M., Silk J., 1999, MNRAS,
submitted
\bi Sunyaev R.~A., Zel$^{\prime}$dovich Ya.~B., 1972, Comm. Astrophys. Space Phys., 4, 173
\bi Tozzi P., Governato F., in : The Young Universe: Galaxy formation at Intermediate and High Redshifts,
ed: Fontana A., Giallongo E., ASP Conference Series, vol 146, 1998, 461 
\bi Vittorio N., Colafranseco S., Mazzotta P., Rephaeli Y., 1997 in : Microwave 
Background Anisotropies, Proceedings of XIV Moriond Astrophysics Meetings, ed: Bouchet
F.,
Gispert R., Guiderdoni B., Jean Tran Thanh Van
\bi Weinberg S., 1972, Gravitation and Cosmology., John Wiley \& Sons Inc., New York
\bi White S.~D.~M., Navarro J.~F., Evrard A.~E., Frenk C.~S., 1993, Nature, 366,
 429

\eeb
\end{document}